\documentclass[conference]{IEEEtran}
\IEEEoverridecommandlockouts
\usepackage{cite}
\usepackage{amsmath,amssymb,amsfonts}
\usepackage{algorithmic}
\usepackage{graphicx}
\usepackage{textcomp}
\usepackage{xcolor}
\usepackage{cite}
\usepackage{amsmath,amssymb,amsfonts}
\usepackage{algorithmic}
\usepackage{graphicx}
\usepackage{textcomp}
\usepackage{xcolor}
\usepackage{ulem}
\usepackage{booktabs}
\usepackage{ulem}
\usepackage{balance}
\usepackage{hyperref}
\usepackage{multirow}
\def\BibTeX{{\rm B\kern-.05em{\sc i\kern-.025em b}\kern-.08em
    T\kern-.1667em\lower.7ex\hbox{E}\kern-.125emX}}
\begin{document}

\title{A Synthetic-to-Real Dehazing Method based on Domain Unification}


\author{
Zhiqiang Yuan$^{1}$ \qquad Jie Zhou$^{1}$  \qquad Jinchao Zhang$^{1, \star}$
\thanks{$^{\star}$ Jinchao Zhang is the corresponding author}
\\Pattern Recognition Center, WeChat AI, Tencent$^{1}$ 
}
\maketitle

\begin{abstract}
Due to distribution shift, the performance of deep learning-based method for image dehazing is adversely affected when applied to real-world hazy images.
In this paper, we find that such deviation in dehazing task between real and synthetic domains may come from the imperfect collection of clean data.
Owing to the complexity of the scene and the effect of depth, the collected clean data cannot strictly meet the ideal conditions, which makes the atmospheric physics model in the real domain inconsistent with that in the synthetic domain.
For this reason, we come up with a  synthetic-to-real dehazing method based on domain unification, which attempts to unify the relationship between the real and synthetic domain, thus to let the dehazing model more in line with the actual situation.
Extensive experiments qualitatively and quantitatively demonstrate that the proposed dehazing method significantly outperforms state-of-the-art methods on real-world images.
\end{abstract}


\section{Introduction}
Haze is a common natural phenomenon that causes color distortion, low contrast, and blur in images, which further degrades the performance of subsequent tasks such as object detection\cite{hu2023content}\cite{hu2024hu2024lf}, semantic segmentation\cite{kirillov2023segment}\cite{zhou2023weakly} and others \cite{walkvlm}\cite{yuan2022exploring}.
Therefore, designing a dehazing network that performs well in real scenarios has become one of the research hotspots in recent years.
The hazing process is usually formulated by the well-known atmospheric scattering model (ASM) \cite{nayar1999vision}: 
\begin{equation}
  I = J e^{-\beta z} + A(1-e^{-\beta z})
  \label{one}
\end{equation}

\noindent
where $I$ and $J$ are the hazy and haze-free image respectively. $A$ denotes the global atmospheric light. $e^{-\beta z}$ is the transmission map, in which $z$ is the scene depth and the scattering coefficient $\beta$ represents the degree of haze.

With the development of deep learning, researchers attempt to utilize deep networks to directly predict the values of $J$,$A$, and $T$ in ASM, and obtain robust performance by making them satisfy physical priors\cite{shao2020domain}\cite{PSD}.
Due to the lack of paired data in real scenarios, researchers have to add haze to the collected clean images by the ASM model, thereby using synthetic paired data to optimize the dehazing model\cite{dong2020multi}\cite{qin2020ffa}.
However, when such models are directly transferred to real scenes, they usually suffer from different degrees of performance degradation due to inconsistent domain\cite{liu2021synthetic}\cite{wang2024frequency}.

\begin{figure}[!ht]

\centering
\includegraphics [width=2.3in]{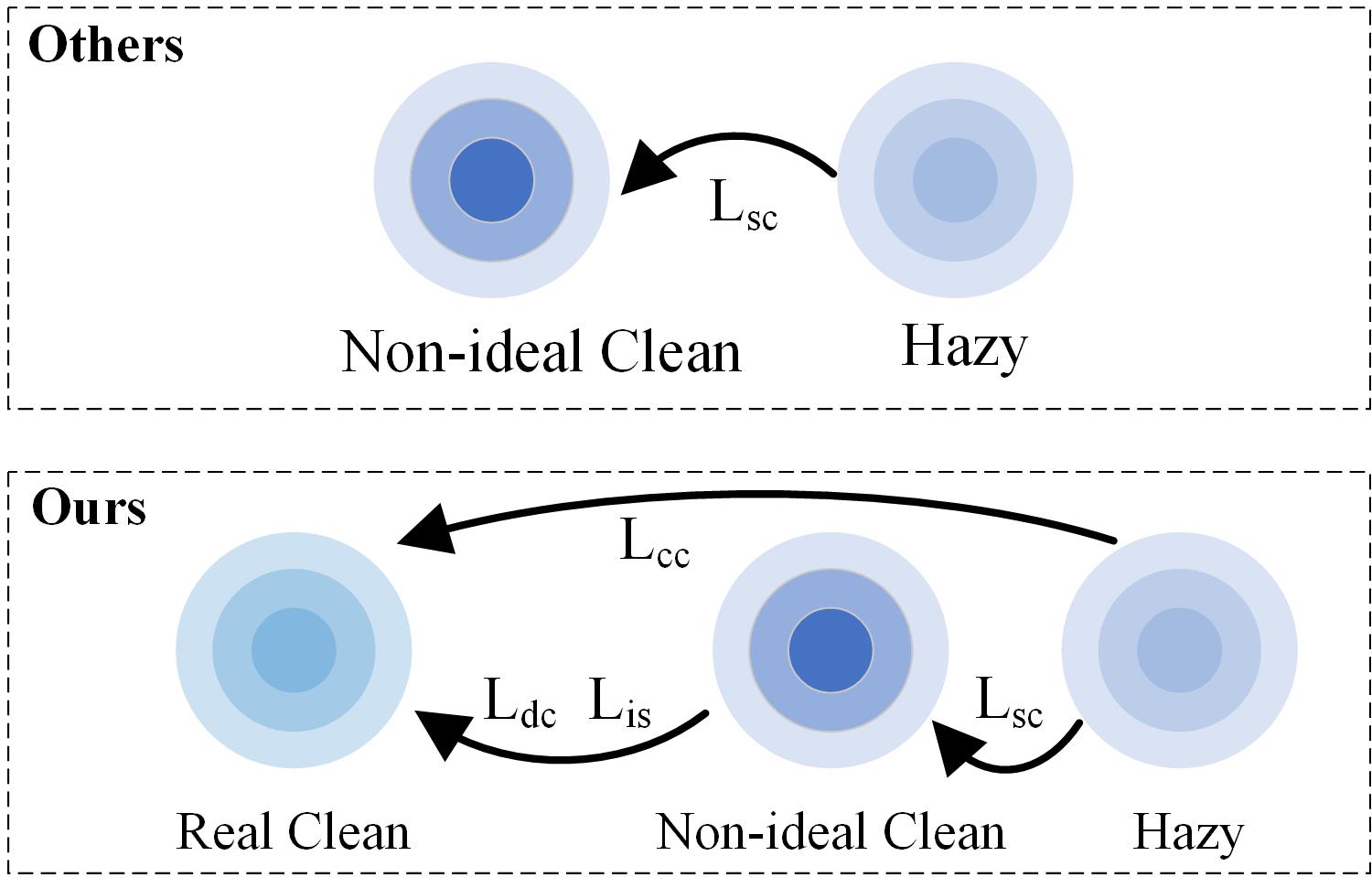}
\vspace{-10px} 
\caption{
Traditional methods typically use paired data for model training based on atmospheric scattering models.
In this paper, we argue that there are non-ideal clean images in the paired data, which will cause the model's inference results to deviate from the real scene.
To this end, we designed a loss committee to constrain multiple aspects so as to generate real clear images in real world.
}
\vspace{-20px} 
\label{task}
\end{figure}

In this paper, we argue that the performance degradation during domain migration may come from the non-ideal data collection.
When the synthetic hazy image is directly hazed using non-ideal clean data, the generated hazy image and the clean image do not strictly hold the atmospheric scattering model.
Therefore, models trained using such synthetic data are not ideal when transferred to real data.
For this reason, we derive the equation between synthetic data in non-ideal situations and clean data in ideal scenarios, and come up with a  synthetic-to-real dehazing method based on domain unification by exploiting the actual transformation between two domains.
The proposed method implicitly models the relationship between real clean images and synthetic hazy images by a loss committee, which constrains the real clean image $J$ from the perspective of intra-domain and inter-domain according to the actual ASM model.
Extensive experiments have demonstrated our method can greatly outperform other SoTA methods on the real-world dehazing task.
The main contributions of our work are as follows:
\vspace{-4px}
\begin{itemize}

\item We point out the dehazing deviation between the real and synthetic domains, which may be caused by non-ideal clean data collection.

\item We come up with a  synthetic-to-real dehazing model based on domain unification, which constrains the dehazing results from both intra-domain and inter-domain perspectives through implicit supervision.

\item Our method performs well on the synthetic-to-real dehazing task, ahead of other state-of-the-art methods such as FCDM and DEA-Net in multiple real-world datasets.

\end{itemize}

\vspace{-5px}
\section{Related Works}

\subsection{Single Image Dehazing}

Existing dehazing algorithms can be broadly categorized into \textit{prior-based methods} and \textit{learning-based methods}\cite{cai2016dehazenet}\cite{singh2019comprehensive}\cite{liu2023visual}. 
\textit{Prior-based methods} refer to  estimating the transmission map using the static properties of a clean image and further integrating the atmospheric scattering model to recover haze-free images\cite{retinex}\cite{fattal2008single}. 
Tang $et \ al.$ \cite{tang2014investigating} developed a regression framework to determine the optimal prior combination for dehazing and introduced haze-related priors. 
Fattal \cite{fattal2014dehazing} derived a local formation model that explains the color-lines in hazy scene and leverage it to recover the scene transmission. 
The \textit{learning-based method} is in a data-driven manner and uses deep neural networks to perform the dehazing process, including discriminative \cite{das2022gca, liu2021synthetic} and generative methods\cite{engin2018cycle, wang2024frequency}.
Early research often used deep neural networks to predict intermediate transmission maps and remove haze based on ASM\cite{zhang2018densely, zhao2017deep}. 
Furthermore, many studies adopted a fully end-to-end approach, training deep neural networks to approximate ground truth haze-free images\cite{dong2020multi}\cite{qin2020ffa}.
Due to the limitations of paired data in dehazing, researchers have begun to focus on using generative networks, such as GANS \cite{Enhanced Pix2pix Dehazing}\cite{dong2020fd} and diffusion models\cite{he2024diffusion}\cite{guo2024multi}, for dehazing.
Although these methods can perform dehazing well in the same domain, the deviation between synthetic and real data cause the models trained on synthetic data perform poorly in real scenes\cite{wang2024frequency}\cite{liu2021synthetic}.

\subsection{Synthetic-to-Real Dehazing}

As mentioned above, although single image dehazing methods achieve well results on synthetic data, performance degradation usually occurs when generalized to real scenes.
To better address the problem of dehazing in real scene, many works \cite{shao2020domain, PSD} are carried out within a paradigm of semi-supervised learning that assumes the availability of some real images. 
Chen $et \ al.$ \cite{PSD} proposed a principled synthetic-to-real dehazing framework to enhance the generalization performance of ASM-based dehazing. 
Domain adaptation and generalization techniques are usually utilized to bridge the gap between synthetic and real data. 
Wang $et \ al.$ \cite{wang2024frequency} attempted to improve the generalization ability of the diffusion-based dehaze model and proposed a frequency compensation module to enhance the learning of high-frequency patterns in the denoising process.
However, the above methods do not consider the problems in the data collection process, which may leads to a certain gap between the synthetic and real domain.

\section{Methods}

In this section, we first describe the problems that imperfect data collection poses to current synthetic-to-real dehazing methods, and then propose a loss committee based on domain unification to implicitly estimate the distribution of clean images in real scenes.

\subsection{Dehaze in Non-ideal Situation}
\label{3.1}

As a stable prior, the atmospheric physics model (ASM) becomes the core theory in dehazing\cite{nayar1999vision}.
Specifically, in real scene, given an observed hazy image $I_c$ and the haze-free image $J$, the relation of them can be expressed as:
\begin{equation}
I_c = e^{-\beta_c z}J + (1-e^{-\beta_c z})A_c
\label{2}
\end{equation}
where $A_c$ and $\beta_c$ are the atmospheric light and scattering coefficient when the clean image was photographed, respectively. 
For ideal haze-free air conditions, the size of the suspended particles in the air is approximately zero, so that $\beta$ tends to zero.
Further, $e^{-\beta_c z}$ tends to one, causing $I_c$ and $J$ to be equivalent at this time.

Due to the lack of paired data in natural scenes, researchers have to use synthetic data to optimize the dehaze model.
During data synthesis, the hazy image $I_h$ is still obtained from an photographed clean image $I_c$ by using ASM model\cite{RESIDE}\cite{zhang2021learning}:

\begin{equation}
I_h = e^{-\beta_h z}I_c + (1-e^{-\beta_h z})A_h
\label{3}
\end{equation}
where $\beta_h$ and $A_h$ are the parameters during synthesis.
Noted that in Equation (3), $I_c$ is used to represent the clean image of synthetic data, because when the air condition is fine, $I_c$ is equivalent to $J$.


\begin{figure}[!ht]

\centering
\vspace{-5px}
\includegraphics [width=3.5in]{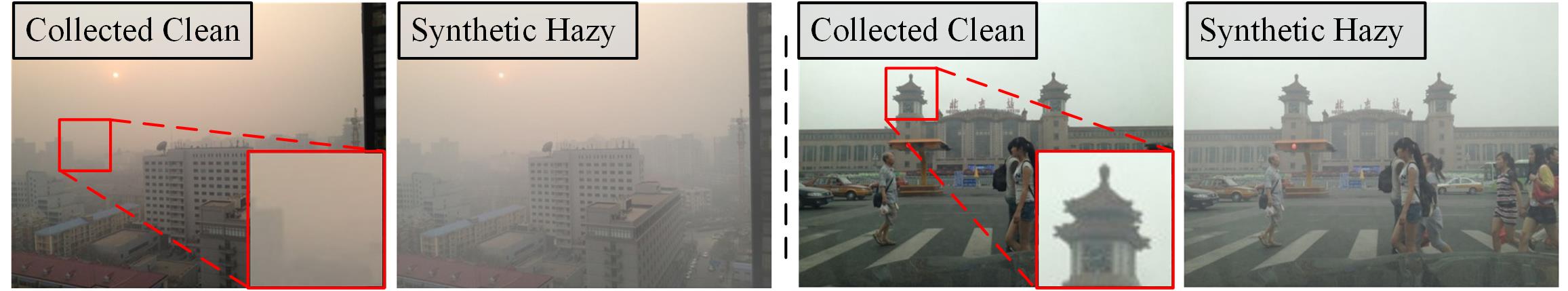}
\vspace{-5px}
\caption{
Data visualization in non-ideal situations.
The collected clean data still contains a lot of haze, especially in areas with greater scene depth.
}
\label{problem}
\vspace{-7px}
\end{figure}

However, as shown in Figure \ref{problem}, although the synthetic process regards the substantially haze-free image as clean image, such conditions are still not ideal due to the complexity of the scene and the effect of depth.
For most synthetic samples, $I_c$ still contains suspended particles in different degrees, which makes $I_c$ and $J$ not equivalent.
In this case, by integrating Equations (\ref{2}) and (\ref{3}), the following formula can be established:

\noindent
\begin{small}
\begin{equation}
\label{J_with_Ih}
\begin{aligned}
&I_h = e^{-\beta_h z}I_c + (1-e^{-\beta_h z})A_h \\
&= e^{-\beta_h z}(e^{-\beta_c z}J + (1-e^{-\beta_c z})A_c) + (1-e^{-\beta_h z})A_h \\
&= e^{-(\beta_h+\beta_c) z}J + e^{-\beta_h z}(1-e^{-\beta_c z})A_c + (1-e^{-\beta_h z})A_h
\end{aligned}
\end{equation}
\end{small}

\noindent
It can be seen that in Equation (\ref{J_with_Ih}), the synthesized hazy image $I_h$ and the real clean image $J$ do not strictly satisfy the ASM model.
The distribution of the synthesized data is inconsistent with the distribution of the real scene data, which is the direct reason of performance degradation when the trained model is tested on the real data.
In order to keep the test and real data distribution consistent, we attempt to dehaze by the real correspondence in Equation (\ref{J_with_Ih}) between synthetic data $I_h$ and real clean image $J$, aiming to let the model adapt to the data distribution in the real scene.


\subsection{Dehazing with Domain Unification }
In this section, we unify model training for synthetic and real domains under non-ideal conditions.
Given a synthetic haze image $I_h$, traditional methods \cite{liu2021synthetic}\cite{Domain Adaptation for}\cite{PSD} directly predict the transmission map $T_h$, the atmospheric light $A_h$ and the clean image $I_h$, and perform image restoration through ASM.
Unlike traditional methods, we adopt a two-stage dehazing approach.
Firstly, we directly let the model predict $\beta_h$ and $z$, which are the variables related to $T_h$:
\begin{equation}
\begin{aligned}
I_c, A_h, \beta_h, z_h &= Dehaze(I_h) \\
\end{aligned}
\label{5}
\end{equation}
where $Dehaze$ denotes the dehazing model.
Secondly, as discussed in Section $\ref{3.1}$, in most synthesis scenarios, $I_c$ still contains suspended particles in different degrees.
Therefore, for $I_c$, we can still use dehaze network to obtain relational parameters:
\begin{equation}
\begin{aligned}
J, A_c, \beta_c, z_c &= Dehaze(I_c) \\
\end{aligned}
\label{6}
\end{equation}
After obtaining these variables, we can design relevant loss committee to constrain them during training:

\noindent
\textbf{Synthesis-domain Consistency Loss.}
Consistent with traditional methods, for paired data in the synthetic domain, we use ASM to constrain the relationship between $I_c$ and $I_h$:

\begin{equation}
\mathcal{L}_{sc} = L_1(
e^{-\beta_c z_c} I_c + (1-e^{-\beta_c z_c})A_c, \quad I_h)
\end{equation}
where $L_1$ is the L1 loss.

\noindent
\textbf{Cross-domain Consistency Loss.}
In addition to the constraints of the synthetic domain, given the relationship between the real clean image and the synthetic hazy image in Equation (\ref{J_with_Ih}), we use it to constrain the estimated real clean image $J$:
\begin{equation}
\begin{aligned}
\mathcal{L}_{cc} = & L_1(e^{-(\beta_h+\beta_c) z}J + e^{-\beta_h z}(1-e^{-\beta_c z})A_c 
\\ & + (1-e^{-\beta_h z})A_h, I_h)
\end{aligned}
\end{equation}
In the implementation, since $I_c$ and $J$ have the same scene depth, we treat both $z_c$ and $z_h$ as $z$.

\noindent
\textbf{Depth Consistency Loss.}
Since the depth of the same image in different states is constant, it is more intuitive to directly constrain the distance between $z_h$ and $z_c$.
In addition, considering the importance of depth in the dehazing model, we let the model distill depth prediction of DepthAnything\cite{yang2024depth}.
Therefore, the depth consistency loss is defined as:
\begin{equation}
\mathcal{L}_{dc} = L_1(z_h, Depth(I_c)) +  L_1(z_c, Depth(I_c))
\end{equation}
in which $Depth$ is the DepthAnything for depth prediction.

\noindent
\textbf{Implicit Supervision Loss.}
Since there is no ideal clean image in real scenes, $J$ in Equation (\ref{J_with_Ih}) is used as an implicit predictor.
Without additional supervision, such implicit variables are difficult to learn, which is also discussed in ablative study.
To construct stronger supervision, we expect $J$ to be consistent with the distribution of $I_c$ in the real scene, so KL divergence is leveraged to narrow the distance between them:
\begin{equation}
\mathcal{L}_{div} = KL( J, I_c )
\end{equation}
where we used a convolutional network to extract features from the image.

After constructing the loss committee, we set different weights $\alpha$ to balance the losses:
\begin{equation}
\mathcal{L} = \alpha_{sc}\mathcal{L}_{sc} +
\alpha_{cc}\mathcal{L}_{cc} +
\alpha_{dc}\mathcal{L}_{dc}+  
\alpha_{is}\mathcal{L}_{is}
\end{equation}

\noindent
The designed loss committee constrains the latent variable $J$ from multiple perspectives by unifying different domains, thereby reducing the domain bias caused by incomplete data collection compared to other methods.

As shown in Equations (\ref{5}) and (\ref{6}), a two-stage dehazing model is used in training.
During the inference process, since the model adapts to the distribution of the clean image $I_c$ in real scene, only one inference is needed to get the dehazed result $J$.

\begin{table*}[!t]
	\centering
  \linespread{0.92}
	\caption{
 Quantitative comparison on real-world datasets. \textbf{Bold} and \uline{underline} indicate the best and the second-best, respectively.
 }
 \resizebox{\textwidth}{!}{\begin{tabular}{c|c|ccccccccccccc}
		\toprule
		Datasets & \multicolumn{1}{c|}{Indicators} & \multicolumn{1}{c}{DCP}  & \multicolumn{1}{c}{FFANet} & \multicolumn{1}{c}{GCANet}  & \multicolumn{1}{c}{MSBDN} & \multicolumn{1}{c}{DF} & \multicolumn{1}{c}{$PSD_{\text{GCA}}$} & \multicolumn{1}{c}{$PSD_{\text{MSBDN}}$} & \multicolumn{1}{c}{D4} & \multicolumn{1}{c}{FCDM} & DEA-Net & Ours \\
		\midrule
		\multirow{5}[2]{*}{I-Haze} & PSNR $\uparrow$ & 16.037      & 16.887  & 16.979   & 16.988  & 16.550  & 14.694 & 14.822 & 16.938 & \uline{17.346} & 16.293 & \textbf{17.971} \\
		& SSIM $\uparrow$   &  0.718    &  0.739  & 0.746  & 0.748  & 0.754  & 0.674 & 0.717 & 0.744 &0.823 & \uline{0.828} & \textbf{0.846}\\
		& CIEDE $\downarrow$    & 25.349  &   28.437  & 23.892   & 24.317  & 12.567  & 28.731  & 29.173 & 13.155 & \uline{11.529} & 12.128 & \textbf{10.371} \\
            & MUSIQ $\uparrow$   & 58.964     & 46.755  & 62.118    & 63.340  & 64.961  & 58.399  & 63.011 & 64.514 & \uline{68.792} & 65.291 & \textbf{69.724} \\
		& NIQE $\downarrow$    & 5.225  &  7.628  & 5.071    & 5.428  & 4.378  & 3.975 & 3.912 & 4.539 & \textbf{3.817} & 3.926 & \uline{3.857} \\
		\midrule
		\multirow{5}[2]{*}{O-Haze} & PSNR $\uparrow$   & 14.622        & 16.279  & 14.555  & 16.562  & 16.340  & 13.774 & 12.676 & 16.922 &17.389 & \uline{17.432} & \textbf{18.237} \\
		& SSIM $\uparrow$    & 0.676   &  0.661  & 0.652  & 0.656   & 0.700  & 0.567 & 0.605  & 0.607 & \uline{0.802} & 0.787 & \textbf{0.813} \\
		& CIEDE $\downarrow$    & 25.956  &  23.952  & 25.069    & 23.284  & 15.897  & 28.480  & 24.031 & 15.195 &\uline{13.583} & 14.901 & \textbf{12.647} \\
            & MUSIQ $\uparrow$ &64.017    & 62.700  & 63.388   & 63.157  & 63.908  & 56.650  & 63.047 & 63.042 &\uline{64.549} & 63.782 & \textbf{64.717} \\
		& NIQE $\downarrow$ & 5.230    &  3.232  & 3.817   & 4.465  & 3.054  & 3.246 & 2.744 & 2.610 &\textbf{2.268} & 2.872 & \uline{2.473} \\
		\bottomrule
	\end{tabular}
}
\label{Reference Metrics}
\end{table*}

\begin{figure*}[!t]
\centering
\includegraphics [width=6.8in]{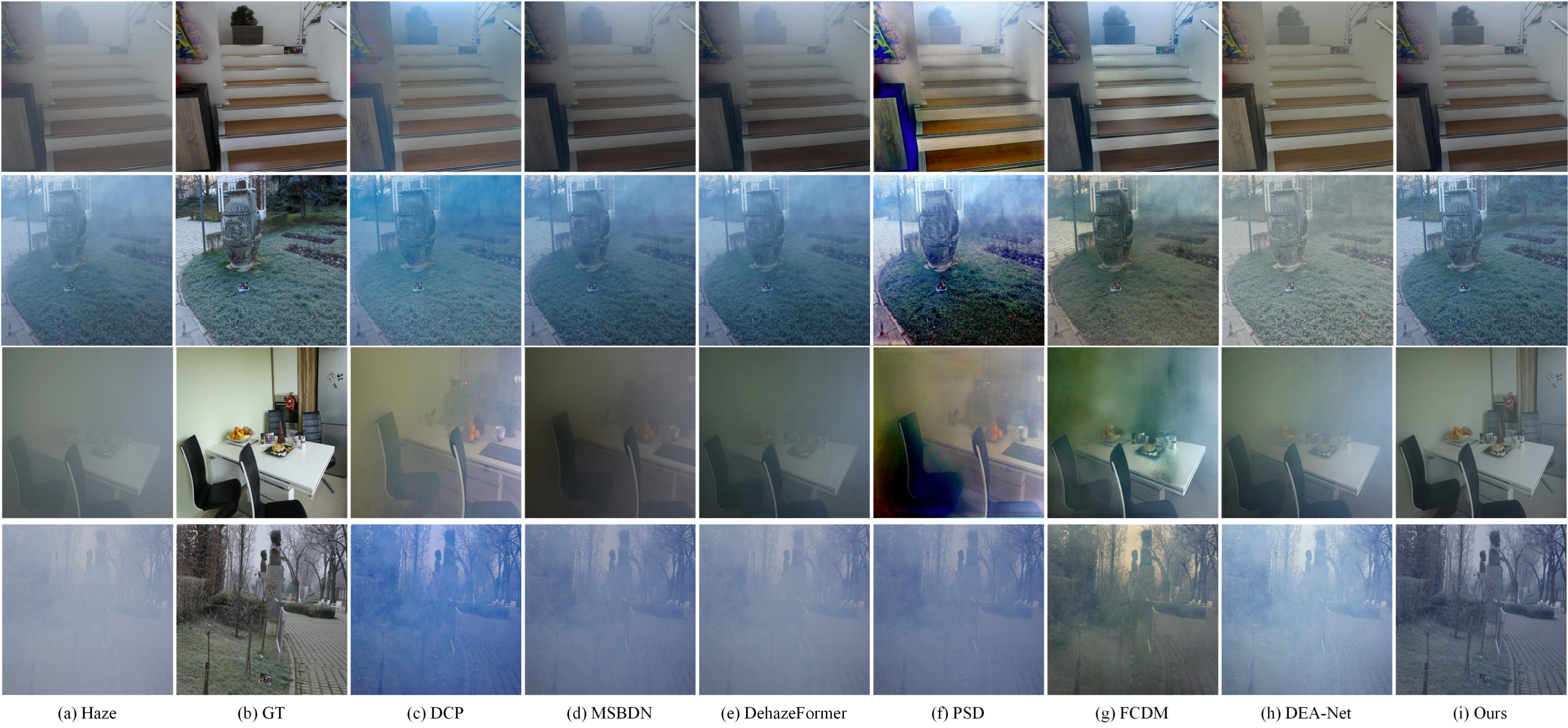}
\vspace{-7px}
\caption{
  Qualitative comparison of dehazed images using different methods.
  From top to bottom are the visualization on the different database. 
}
\label{visual_sota}
\vspace{-10px}
\end{figure*}

\section{Experiments}

This section presents and analyzes the experimental results, which verify the well generalization of the proposed method.

\subsection{Experiment Settings}

\noindent
\textbf{Datasets.}
The RESIDE dataset \cite{RESIDE} is widely recognized in dehazing tasks, encompassing five distinct subcategories: ITS, OTS, SOTS, RTTS, and HSTS. 
The subsets ITS, OTS, and SOTS are synthetic datasets, while RTTS is a real-world dataset.
In our experimental setup, we adhered to the precedent set by \cite{wang2024frequency}, opting for the ITS and OTS datasets as our training datasets. We then employed the trained model to make predictions in real-world scenarios, including I-haze, O-haze, and RTTS, to rigorously assess the model's generalization capabilities.

\noindent
\textbf{Comparison methods.}
We compare our model against current state-of-the-art dehazing algorithms based on conventional methods (DCP \cite{he2010single}), convolutional neural networks (FFANet \cite{qin2020ffa}, GCANet \cite{das2022gca}, MSBDN \cite{dong2020multi}, DEA-Net \cite{chen2024dea}), ViT (DehazeFormer (DF) \cite{song2023vision}) and synthetic-to-real methods (PSD \cite{PSD}, D4 \cite{yang2022self}, FCDM \cite{wang2024frequency}).
For MSBDN, DF, and FCDM, we utilize the provided pre-trained models and for the rest, we use the provided codes to retrain models on the same dataset leveraged in this paper for fair comparison.

\noindent
\textbf{Metrics.}
For traditional dehazing methods, Peak signal to noise (PSNR), structural similarity index (SSIM)  and CIEDE \cite{hore2010image}\cite{CIEDE} 
are often used to evaluate the effectiveness of various methods.
However, since ideal real images are difficult to obtain, it is suboptimal to only utilize pixel-level evaluation metric to obtain a comprehensive evaluation result.
For this reason, we also leverage non-reference metrics  NIQE \cite{NIQE} and MUSIQ\cite{ke2021musiq} to measure how much the dehazed images are similar to natural images in terms of statistical regularities and granularities. 

\noindent
\textbf{Implementation details.}
We use DEA-Net as the backbone and adds the loss committee for training.
The loss weights $\alpha_{sc}$, $\alpha_{cc}$, $\alpha_{dc}$, and $\alpha_{is}$ during training are 0.5, 0.3, 0.05, and 0.1 respectively.
The weights of the loss committee are obtained through Bayesian optimization in AutoML\footnote{https://github.com/sb-ai-lab/LightAutoML}.
We used ResNet-18 \cite{resnet} to extract image representations, thereby calculating the KL loss.
During training, we resize each image to
256$\times$256, and average on the last five epoch weights via early stopping to get the result\cite{hu2025sac}.
Loss $\mathcal{L}_{sc}$ is only used for the first 30 epochs of training, then loss committee is utilized for the next 70 epochs.

\subsection{Compared to Other Methods}

\noindent 
\textbf{Quantitative comparison.}
Our approach demonstrates superior performance over other established methods across different datasets, as illustrated in Table \ref{Reference Metrics}, especially in the referenced metrics.
For non-reference metrics, in terms of the MUSIQ indicator for measuring image granularities, our method is still ahead of FCDM.
In NIQE evaluation, our method is slightly inferior to FCDM. 
The reason may come from the implicit reconstruction of the clean image $J$, while FCDM explicitly fits the distribution of $J$ by leveraging the powerful generation ability of the diffusion model.
Table \ref{RTTS} shows the evaluation results on RTTS, and the conclusions are consistent with the above.

\begin{table}[!ht]
\centering
  \linespread{0.92}
 \scriptsize
 	\caption{Quantitative comparison on RTTS datasets. 
\textbf{Bold} and \uline{underline} indicate the best and the second-best, respectively.}
\resizebox{\linewidth}{!}{
\begin{tabular}{c|ccccccc}
\hline
Indicators & DCP                     & $PSD_{\text{MSBDN}}$  & D4    & FCDM & DEA-Net & Ours        \\ \hline
MUSIQ  $\uparrow$  & 55.079     & 56.712 & 56.809 & \uline{58.218} & 56.924 & \textbf{58.974} \\
NIQE  $\downarrow$  & 3.852    & 3.667 & 4.415 & \textbf{2.968} & 4.437 & \uline{3.117} \\ \hline
\end{tabular}
}
\label{RTTS}
\end{table}

\noindent
\textbf{Qualitative comparison.}
Figure \ref{visual_sota} presents a visual comparison of dehazed images generated by different methods. Our method stands out by delivering high-quality, haze-free images characterized by distinct edges and minimal residual haze. Notably, in regions with substantial scene depth, such as the stair corners and dense woodlands depicted in Figure \ref{visual_sota}, our approach yields notably clearer dehazing outcomes.
Furthermore, our method excels in preserving the original image's color integrity with excellent fidelity output.

\begin{figure}[!ht]
\centering
\includegraphics [width=3.0in]{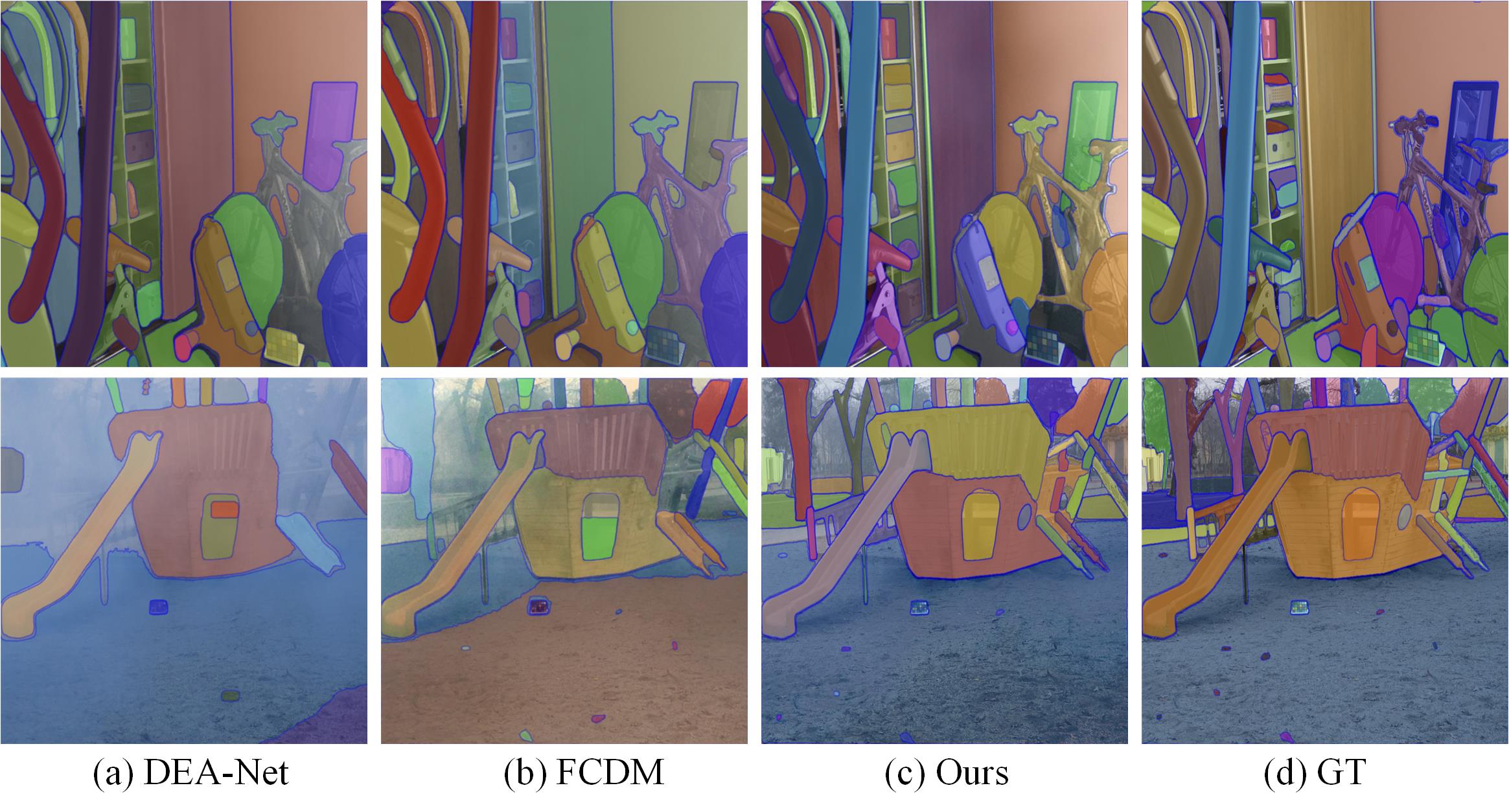}
\caption{
Comparison of segmentation results after dehazing using different methods.
Our results can produce more details after segmentation.
}
\label{sam}
\end{figure}

\begin{figure}[!ht]
\centering
\includegraphics [width=3.2in]{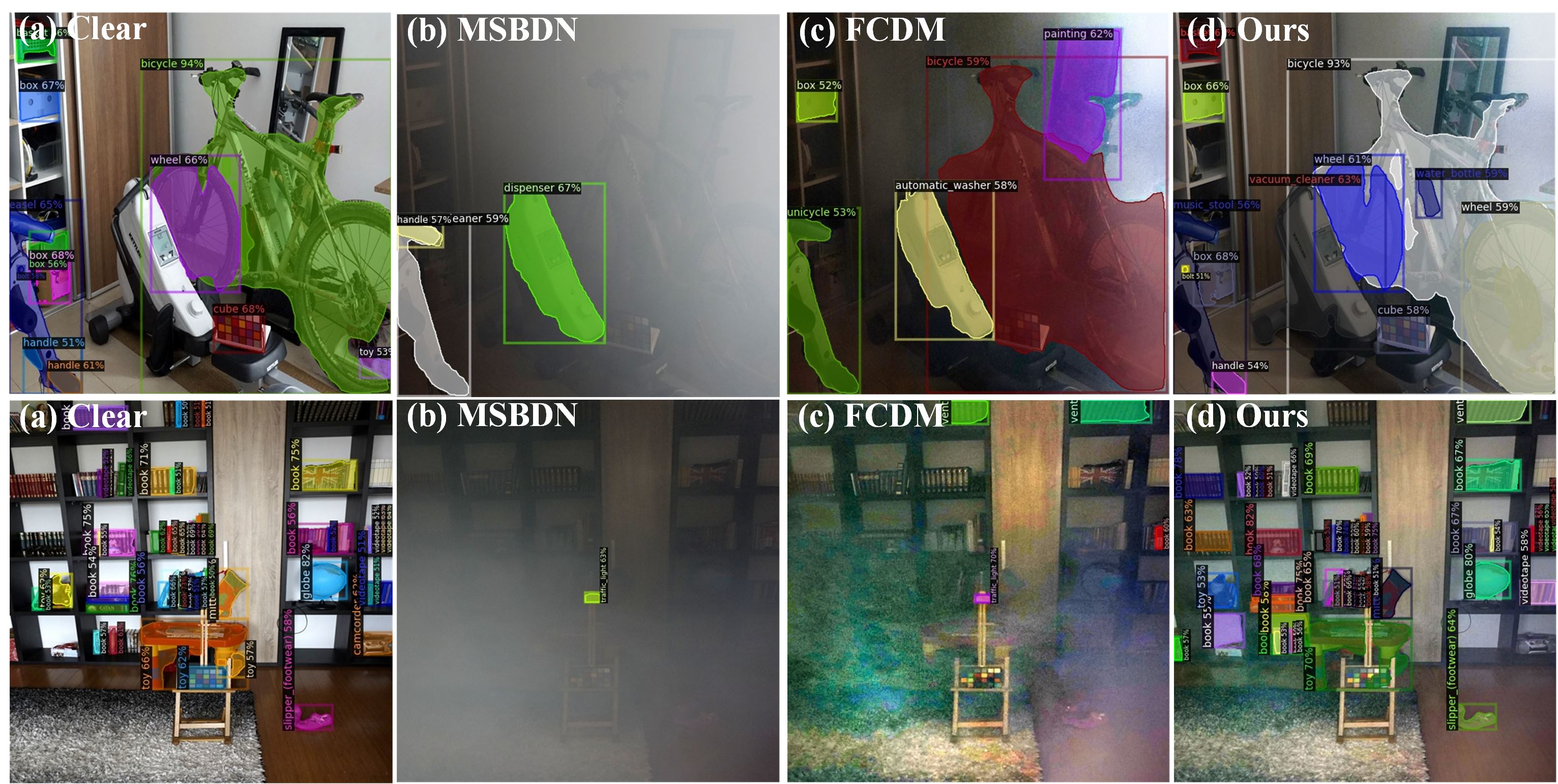}
\caption{
Comparison of detection results after dehazing using different methods. Compared to other methods, our approach can generate more accurate and more confident detection results.
}
\label{fig5}
\end{figure}

\noindent
\textbf{Performance comparison on downstream task.}
The segmentation task is take as a case study to verify the proposed dehazing method on downstream tasks.
We use SAM\cite{kirillov2023segment} to segment the dehazed results of DEA-Net, FCDM, and ours respectively, and make qualitative comparisons with the ground truth.
As shown in Figure \ref{sam}, our segmented results are more fine-grained in areas such as bookshelves and branches, and are more consistent with the ground truth.
Additionally, as shown in Figure \ref{fig5}, we also compared the detection results after dehazing with different models. Our model was able to achieve more confident detection results.

%
\subsection{Ablative Study}

\noindent 
\textbf{Quantitative ablation.}
This subsection conducts ablation experiments to essay the role of individual loss.
As shown in Table \ref{abalation}, we set up four sets of comparative experiments.
\begin{itemize}
    \item $m_1$: Using only synthesis-domain consistency loss $\mathcal{L}_{sc}$ to constrain the model, consistent with traditional methods.
    \item $m_2$: Cross-domain consistency loss $\mathcal{L}_{cc}$ is added to $m_1$, but the implicit clean $J$ is not constrained by KL loss.
    \item $m_3$: 
    Implicit supervision loss $\mathcal{L}_{is}$ is added to $m_2$. 
    
    \item $m_4$: $\mathcal{L}_{dc}$ and $\mathcal{L}_{is}$ are used to constrain $J$ on the $m_2$ model.
\end{itemize}
$(m_1,m_2)$ aims to explore the performance changes when only adding cross-domain consistent loss, $(m_2, m_3)$ explores the importance of constraining the implicit variable $J$, and $(m_3, m_4)$ explores the importance of depth.

\begin{table}[!ht]
\centering
\caption{Ablation study of different losses on I-Haze datasets.  \\
* denotes unstable result.
}
\begin{tabular}{c|cccc|cc}
\hline
   & $\mathcal{L}_{sc}$                       & $\mathcal{L}_{cc}$                      & $\mathcal{L}_{dc}$                       & $\mathcal{L}_{is}$                       & PSNR $\uparrow$ & SSIM $\uparrow$ \\ \hline
$m_1$ & \checkmark &                           &                           &                           &    16.293  &    0.828  \\
$m_2^*$ & \checkmark & \checkmark &                           &                           &   15.736   &  0.817    \\
$m_3$ & \checkmark & \checkmark &                           &  \checkmark                         &   16.814   &  0.833    \\
$m_4$ & \checkmark & \checkmark & \checkmark & \checkmark &  \textbf{17.971}    &  \textbf{0.846}    \\ \hline
\end{tabular}
\label{abalation}
\end{table}

Table \ref{abalation} delineates the results of our model under various configurations. 
When the model $m_1$ is augmented solely with loss $\mathcal{L}_{cc}$, there is a marginal decline in performance. 
This decrement can be attributed to the inadequate constraints imposed on the latent variable $J$.
In this case, the result of $m_2$ is unstable.
$m_3$ imposes a KL constraint on $J$, making the change of $J$ more stable during training.
Upon the incorporation of loss $\mathcal{L}_{dc}$ and $\mathcal{L}_{is}$, which serve to further refine the constraints on depth $z$ and the reconstructed image $J$, the model $m_4$ achieves optimal performance.

\begin{figure}[!ht]
\centering
\includegraphics [width=3.0in]{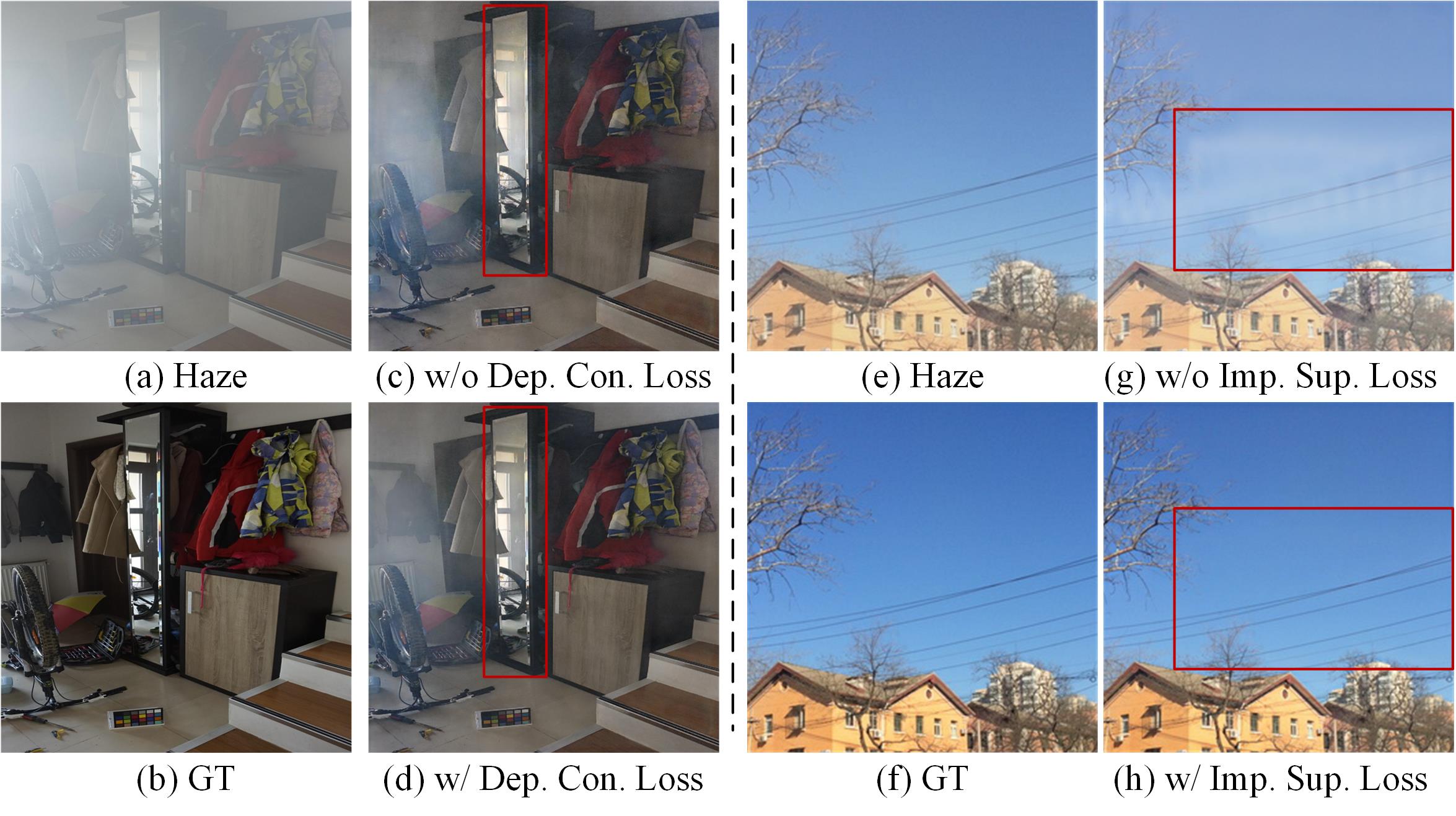}
\caption{
Left and right show the ablation of depth consistency loss and implicit supervision loss, respectively, where the red box area is especially obvious. 
}
\label{s1_s2}
\end{figure}

\noindent
\textbf{Qualitative ablation.}
Figure \ref{s1_s2} (left) presents the qualitative ablation study of 
$\mathcal{L}_{dc}$, emphasizing the significant variation observed in the deeper regions.
Figure \ref{s1_s2} (right) depicts the scenario in the absence of $L_{is}$ as a constraint, resulting in the emergence of textures with varying degrees of granularity within the image.

\subsection{Generalization Capability}

\begin{table}[!ht]
\centering
\vspace{-10px}
  \linespread{0.92}
 	\caption{Comparison of model performance  under \\ different training methods on I-Haze dataset.}
\begin{tabular}{c|ccccccc}
\hline
Training methods & FFANet                     & MSBDN  & DF          \\ \hline
PSNR  $\uparrow$  & 16.887     & 16.988 & 16.550  \\
SSIM  $\uparrow$  & \textbf{17.261}    & \textbf{17.294} & \textbf{16.806}  \\ \hline
\end{tabular}
\label{lgenelize}%
\end{table}

In this subsection, we use this method to train other networks to determine the generalization performance.
We tested FFANet, MSBDN, and DehazeFormer, which are regression-based trained models, and the results are shown in Table \ref{lgenelize}.
Compared to training methods based solely on regression, the training method based on domain unification has enabled different models to achieve better results on real-world datasets, which greatly verifies the generalizability of the proposed method.

\section{Strengths and Limitations}

This paper theoretically validates the bias generated when the ideal atmospheric scattering model is applied to non-ideal state data.
In order to enable the model to cope with such biases, we use pseudo-clear images collected in real scenes as a medium to predict real clear images under ideal conditions that are difficult to collect.
In this process, to maximize the preservation of the features of the real clear image, we designed a loss committee to align the clear images under ideal and non-ideal conditions from different feature levels.
However, we found that reconstructing real clear is non-trivial, and the model is prone to collapse during training, which may require adding more constraints at the feature or pixel level to avoid this situation.

\section{Conclusion}

In this paper, we point out the deviation between real and synthetic domains in the dehazing task due to imperfect collection of clean data.
To address this problem, we propose a synthetic-to-real  dehazing model based on the domain unification and achieve SoTA results on real-world dehazing tasks.

\bibliographystyle{IEEEbib}

\end{document}